\documentclass[a4paper]{article}

\usepackage{icrc2013}
\usepackage[english]{babel}

\title{On the description of the GCR intensity in the last three solar minima}

\shorttitle{GCR intensity in the last three solar minima}

\authors{
Kalinin M.S., Bazilevskaya G.A., Krainev M.B., Svirzhevskaya A.K., Svirzhevsky N.S.
}

\afiliations{
Lebedev Physical Institute, Russian Academy of Sciences, Moscow, Russia
}

\email{mkrainev46@mail.ru}

\abstract{We discuss the main characteristic features in the heliospheric parameters important for the GCR intensity modulation for the last three solar minima (1986--1987, 1996--1997 and 2008--2009). The model for the GCR intensity modulation is considered and the set of the model parameters is chosen which allows the description of the observed GCR intensity distributions at the moments of the maximum GCR intensity in two solar minima (1987 and 1997) normal for the second half of the last century. Then we try to describe with the above model and set of parameters the unusually soft GCR energy spectra at the moments of the maximum GCR intensity in the last solar minimum between cycles 23 and 24 (2009). Our main conclusion is that the most simple way to do so is to reduce the size of the modulation region and, probably, change the rigidity dependence of the diffusion coefficient. The change of both parameters is substantiated by the observations of the solar wind and heliospheric magnetic field.}


\begin{document}
\maketitle

%Begin a section.
\section{Introduction}
When compared with the solar cycles (SCs) of the second half of the 20--th century the minimum 23/24 between the SC 23 and SC 24 is rather strange. First of all it concerns the record-setting low strength of the solar and heliospheric magnetic fields (HMF) and the high GCR intensity \cite{Sheeley_ASPCS_428_3_2010}--\cite{Gushchina_etal_JP:CS_409_012169_2013}, but some other details are also unusual \cite{Svirzhevsky_etal_ICRC31_icrc1105_2009,Bazilevskaya_etal_ASR_49_784-790_2012}. Some efforts were made to understand these effects \cite{Moraal_Stoker_JGR_115_A12109_2010,Krainev_Kalinin_Bull_RAS_Physics_75_7862_2011,Gerasimova_etal_ICRC32_11_179,
Kota_Jokipii_ICRC32_11_12-14_2011}, however there are a lot of questions still.

This paper deals with the GCR intensity distributions and heliospheric characteristics in the moments of maximum GCR intensity in the last three solar minima (1987, 1997 and 2009), while the time and energy behavior of the GCR intensity in the periods of low solar activity around these minima are considered in the accompanying papers \cite{Bazilevskaya_etal_ICRC33_0274_2013} (observations) and \cite{Krainev_etal_ICRC33_0305_2013} (modeling). First we discuss the behavior of the main heliospheric parameters important for the GCR intensity modulation. Then, using a very simple model, we try to reproduce the observed GCR intensity distributions (the energy spectrum and latitudinal and radial gradients) at the moments of the maximum GCR intensity in two solar minima (1987 and 1997) which we consider as normal for the second half of the last century.  This task fixes the set of the constant model parameters. Then we try to describe with the above model and set of parameters the unusually soft GCR energy spectra at the moments of the maximum GCR intensity in the last unusual solar minimum between cycles 23 and 24 (2009).

\section{The GCR intensity and heliospheric parameters in 1980--2013}
\begin{figure}[t]
  \centering
  \includegraphics[width=0.5\textwidth]{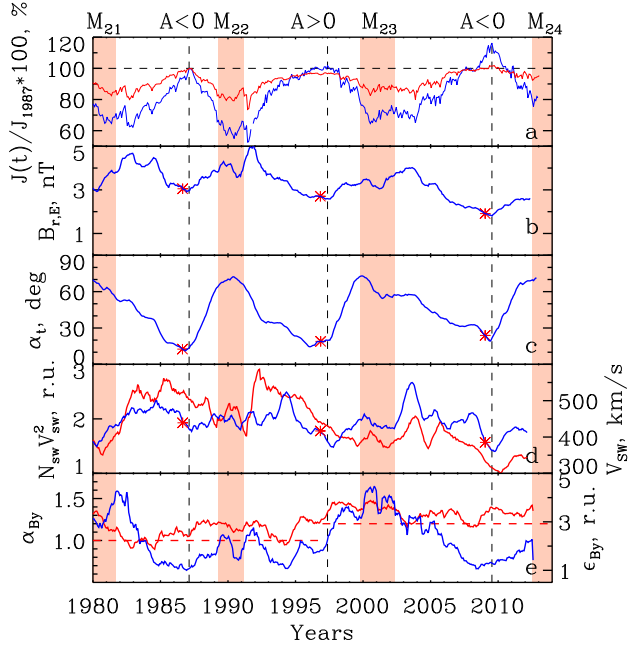}
  \caption{The GCR intensity and heliospheric characteristics in 1980--2013. The periods of solar maxima are shaded. The times of maximum of GCR intensity are shown by the dashed vertical lines, while the red stars indicate the level of some characteristics averaged for one year before these moments. In the panels: (a) the normalized to 100\% for 1987.2 monthly averaged count rates of the RBM omnidirectional detector in Pfotzer maximum above Murmansk (blue) and neutron monitor Moscow (red); (b) the radial component of the regular HMF $B_r$; (c) the HCS tilt $\alpha_t$; (d) the velocity (blue) and dynamic pressure (red) of the solar wind; (e) the energy density (blue) and the index (red) of the inhomogeneity spectrum of the HMF component $B_y$ normal to the average field.}
  \label{fig1}
 \end{figure}

In Fig. \ref{fig1} the time profiles are shown of the GCR intensity and the main heliospheric characteristics which, as we believe, are responsible for the GCR strange features in the 2000s when compared with the 1980--1990s. The GCR behavior is illustrated using the data of the regular balloon monitoring (RBM), the  monthly averaged count rates of the RBM omnidirectional detector in Pfotzer maximum above Murmansk, $N_m^{Mu}$,  $R_c=0.6$ GV, and Moscow, $N_m^{Mo}$,  $R_c=2.4$ GV \cite{Bazilevskaya_Svirzhevskaya_SSR_85_431-521_1998,Stozhkov_etal_Preprint_LPI_14_2007}, and neutron monitor (NM, Moscow, \cite{NM_Moscow}, the effective energy $T_{eff}\approx 15$ GeV). The heliospheric characteristics near the Earth \cite{OMNI_Site} and the tilt $\alpha_t$ of the heliospheric  current sheet (HCS; \cite{WSO_Site}, classic) are yearly smoothed.

In Fig. \ref{fig1} one can see the puzzling difference in the behavior of the low-- and high--energy GCR intensities around the solar minimum 23/24 (2009) when compared with the previous two minima 21/22 (1987) and 22/23 (1997). In 1997 both intensities attained approximately the same levels they had in 1987 although there was small difference because of the magnetic cycle. In 2009 the low--energy intensity jumped up by $\approx 15$ \% while the high--energy one increased by only $\approx 2-3$ \% (or even $\approx 0$ \% for NM Tsumeb \cite{bib:izmiran}).

It is useful to compare the main heliospheric factors important for the GCR modulation averaged for one year before the maxima of the GCR intensity for solar minima 21/22, 22/23 and 23/24. The absolute values of the radial component $B_r$  that strongly influences both the diffusion and drift of charged particles are 3.05, 2.71, 1.91 nT. The HCS tilt also important for the drift of GCRs is somewhat larger in the 2000s than in 1980s and 1990s (12.56, 18.97, 23.85 degs for solar minima 21/22, 22/23 and 23/24, respectively). The solar wind velocity convecting GCR out of the heliosphere also changes in phase with $B_r$ (439, 414, 385 km/s). Note that all of these heliospheric characteristics show gradual change from solar minimum 21/22 through 22/23 to 23/24.

However, beside the above well known characteristics modulating the GCR intensity there are some factors which changed rather abruptly with the beginning of SC23. First, the dynamic pressure of the solar wind $\propto N_{sw}V_{sw}^{2}$ which could determine the distance to the termination shock changed even greater than $B_r$ (as 2.5, 2.0, 1.1 in relative units) for solar minima 21/22, 22/23 and 23/24, respectively. Consequently the effective size of the GCR modulation region ($r_{max}\propto\sqrt{N_{sw}V_{sw}^2}$) could change, see also \cite{Mewaldt_SSR_online_2013}. Besides, the spectrum of the inhomogeneities for the HMF component normal to the average field changed rather abruptly around 1997 its index $\alpha_{B_y}$ increasing by 20--30 \% \cite{Grigoriev_Starodubtsev_BullRAS_Physics_75_801_2011,Gerasimova_etal_ICRC33_0266_2013}. This could result in the different rigidity dependence of the diffusion coefficients (as $K_\parallel\propto R^{\alpha_R},\alpha_R=2-\alpha_{B_y}$) for solar minimum 23/24 when compared with minima 21/22 and 22/23. Note that  the total energy density of this HMF component showed no pronounced change \cite{Grigoriev_Starodubtsev_BullRAS_Physics_75_801_2011,Gerasimova_etal_ICRC33_0266_2013}.

\section{The model}
The GCR intensity $J(\vec r,T,t)$ in the heliosphere in steady case is described by solving the differential boundary--value problem for the distribution function $U(\vec r,p,t)=J(\vec r,T,t)/p^2$ \cite{Parker_PhysRev_110_1445_1958,Krymsky_GaA_4_977_1964,Jokipii_Levy_Hubbard_ApJ_213_861_1977}:
\begin{equation}
	- \nabla\cdot ({\cal K}\nabla U)+{\vec{V}}^{sw}\nabla U-\frac{\nabla\cdot{\vec{V}}^{sw}}3p\frac{\partial U}{\partial p}+{\vec{V}}^{dr}\nabla U=0 \label{TPE}
\end{equation}

\noindent with the usual boundary conditions at $r=r_{min},r_{max}$ and poles (without termination shock and heliosheath) and the "initial" condition $\left.U\right|_{p=p_{max}}=U_{um}(p_{max})$. Here $p$ and $T$ are the momentum and kinetic energy of the particles; $p_{max}= 150$ GeV/c and $U_{um}$ is the distribution function of the unmodulated GCRs; ${\vec{V}}^{sw}$ and ${\vec{V}}^{dr}$ are the solar wind and GCR drift velocities, respectively. ${\cal K}$ is the diffusion tensor with components along the regular HMF,  $K_\parallel$, and across it in the radial, $K_{\perp r}$, and latitudinal, $K_{\perp \vartheta}$, directions. The general expression for the parallel diffusion coefficient is used, $K_\parallel=K^0_\parallel\cdot 5/B_{hmf}\cdot f(R)$, with $K^0_\parallel$ (in 10$^{21}$ cm$^2$/s) depending on the sign of $A$, the dominating HMF polarity, and $f(R)\propto R^{\alpha_R}$ in some rigidity range, while $K_{\perp r}=\alpha_{\perp r}\cdot K_\parallel$, $K_{\perp \vartheta}=\alpha_{\perp \vartheta}\cdot K_\parallel$. The form of the HMF lines corresponds to the Parker HMF $B_P$ with $B_{r,E}$ as the only parameter, but $B_{hmf}$ is modified according to \cite{Jokipii_Kota_GRL_16_1_1989} (see \cite{Krainev_etal_JP:CS_409_012016_2013}), $B_{hmf}=B_P\cdot (1+r^2cos^2\chi \cdot \delta^2_{JK})^{0.5}$ where $\chi$ is the Parker spiral angle. Both $\alpha_{\perp \vartheta}$ and $\delta_{JK}$ are made high at high latitudes and small at lower latitudes using the hyperbolic tangent latitude dependence as in \cite{Potgieter_etal_SpaceSciRev_97_295_2001}. The solar wind velocity $V^{sw}_r$ also depends on $\vartheta$ in the same way, changing from $V^{sw,eq}_r=V_{sw,E}$ at low latitudes to $V^{sw,pol}_r=800$ km/s for $\vartheta < 90-\alpha_t$ and $\vartheta > 90+\alpha_t$. The simple model for the regular and current sheet drift velocities (as in \cite{Krainev_Kalinin_AIPCP_1216_371-374_2010}) is used. So the set of model parameters $\left\{K^{0+}_\parallel,K^{0-}_\parallel,\alpha_{\perp r},\alpha_{\perp \vartheta}^{pol},\delta_{JK}^{pol}\right\}_c$ is constant, while the changing set of the main heliospheric parameters $\left\{B_{r,E},\alpha_t,V_{sw,E} \right\}_m$, and possibly, additional set $\left\{r_{max},\alpha_R\right\}_a$ describe the change of the GCR intensity with time.

As the unmodulated GCR spectrum we used $J_{um}(T)=$
%\vspace{5cm}
\begin{equation}
\left\{
  \begin{array}{l l}
   \textrm{\hspace{-1mm}}0.685\exp\left\{4.47-0.08(\ln T)^2-2.91\sqrt{T}\right\}&\textrm{\hspace{-3.5mm},$T\le 1$}\\
   \textrm{\hspace{-1mm}}0.685\exp\left\{3.22-2.78\ln T-1.9/T\right\}&\textrm{\hspace{-3.5mm},$T>1$}
  \end{array} \right. \label{Jum}
\end{equation}
\noindent with $T$ in GeV. This unmodulated spectrum is the modification of those used in \cite{Langer_Potgieter_JGR_109_1-12_2004,Potgieter_etal_Apxiv_1302.1284_2013} to describe the PAMELA proton spectrum \cite{Adriani_etal_ApJ_765_2_91_2013} at GeV energies.

\section{The solar minima 21/22 and 22/23}
\begin{figure}[h]
  \centering
  \includegraphics[width=0.5\textwidth]{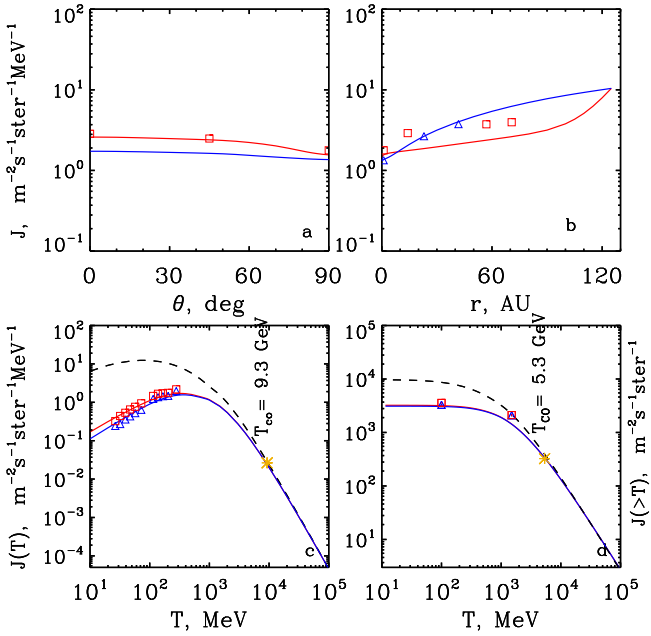}
  \caption{The observed and calculated space and energy distributions of the GCR protons during solar minima 21/22 and 22/23. The red squares and  curves are for the observed  \cite{McDonald_SpaceScience_ser_ISSI_33_1998} and calculated intensity for $A>0$, while the blue triangles and curves are for $A<0$. In the panels: (a) the colatitude profiles for $r=1$ AU, $T=200$ MeV; (b) the radial profiles for $\vartheta=90$ deg, $T=200$ MeV; (c) the differential and (d) integral energy spectra for $r=1$ AU, $\vartheta=90$ deg. The red and blue triangles in panel (d) show the integral GCR intensities estimated using the RBM data. The black dashed lines are for the unmodulated spectra.}
  \label{fig2}
 \end{figure}

First we tried to find the optimal constant set of the model parameters to describe the observed space and energy distributions of the GCR intensity in the solar minima 21/22 (1987) and 22/23 (1997).
Note that choosing the constant set of parameters we tried to get the cross--over of the spectra for $A$--positive and $A$--negative minima at $T_{co}\approx 10$ GeV accounting for the magnetic cycle phase for the RBM and NM data.
In Fig.\ref{fig2} the observed space and energy distributions are compared with the calculated ones using the constant model set  $K^{0+}_\parallel$=15, $K^{0-}_\parallel$=25, $\alpha_{\perp r}$=0.01, $\alpha_{\perp\vartheta}^{pol}$=0.1, $\delta_{JK}^{pol}$=0.39
%$\left\{K^{0+}_\parallel=15,K^{0-}_\parallel=25,\alpha_{\perp r}=0.01,\alpha_{\perp %\vartheta}=0.1,\delta_{JK}=0.39\right\}_c$
and the values of the heliospheric parameters shown by the stars for these solar minima in Fig.\ref{fig1}. The usual values for $r_{max}=125$ AU, $\alpha_R=1$ were used.

As one can see in Fig.\ref{fig2} the calculations with the above model set reproduce the observed space and energy behavior of the GCR intensity with some exceptions. The main drawback is the poor description of the radial profile for $A>0$ solar minimum 22/23 (1997) seen in Fig.\ref{fig2} (b). The second important case of the discrepancy between the observed and calculated GCR intensity is that between the calculated integral intensity and the observed value for $J(>1500)$ MeV for both 21/22 and 22/23 solar minima.

It should be noted that the observed integral intensities $J(>100 \textrm{ MeV})$ and $J(>1500 \textrm{ MeV})$ are calculated using the RBM $N_m^{Mu}$ and $N_m^{Mo}$  and extrapolation to the top of the atmosphere for $N_m^{Mu}$ and $N_m^{Mu-Mo}=N_m^{Mu}-N_m^{Mo}$ using the regressions reported in \cite{Stozhkov_etal_Preprint_LPI_14_2007}. The values for $J(>100 \textrm{ MeV})$ roughly correspond to the unmodulated spectrum used and to the calculated intensities while the estimated values for $J(>1500 \textrm{ MeV})$ are too high.

\section{The unusual solar minimum 23/24}

Then we tried to describe the energy dependence of the maximum GCR intensity in the solar minimum 23/24 (2009). As there are no observations for this period at the intermediate radial distances we do not show the radial dependence of the calculated intensity. We also do not illustrate its latitudinal dependence, although there are the Ulysses data up to the middle of 2009. However, we checked this dependence and state that the latitudinal gradient of the calculated intensity is very small for all cases considered.

So only the energy spectra will be discussed of the intensity calculated for $r=1$ AU, $\vartheta=90$ deg using the same constant model set as was chosen when reproducing the observations in the solar minima 21/22 and 22/23. The set of the main changing modulating parameters $\left\{B_{r,E},\alpha_t,V_{sw,E} \right\}_t$ corresponds to these heliospheric characteristics averaged for one year before the maximum of the GCR intensity in 2009 (shown by the stars in Fig.\ref{fig1}). So for all cases considered only the parameters of the additional set $\left\{r_{max},\alpha_R\right\}$ change.

As observed data for this minimum 23/24 we use the differential spectrum measured for $80<T<43000 \textrm{ MeV}$ by PAMELA in December 2009 \cite{Adriani_etal_ApJ_765_2_91_2013} and the integral intensities  $J(>100 \textrm{ MeV})$ and $J(>1500 \textrm{ MeV})$ estimated from the RBM experiment using the procedure already discussed.
Beside the absolute differential and integral GCR spectra observed and calculated for solar minimum 23/24, we consider
the relative spectra, the ratios of the spectra for the minima 23/24 (2009) and 21/22 (1987) when the HMF polarity was the same ($A<0$). As spectral data for 1987 we use the differential proton spectrum $J(80<T$, MeV $<230)$ measured by IMP8/GME in 01--05.1987 \cite{McDonald_SpaceScience_ser_ISSI_33_1998} and again the integral intensities estimated from the RBM experiment. At high energies the neutron monitor data are used for middle latitude (Moscow, the effective energy $T_{eff}\approx 15$ GeV) and also for the low latitudes (Tsumeb, $T_{eff}\approx 20$ GeV).

 \begin{figure}[h]
  \centering
  \includegraphics[width=0.5\textwidth]{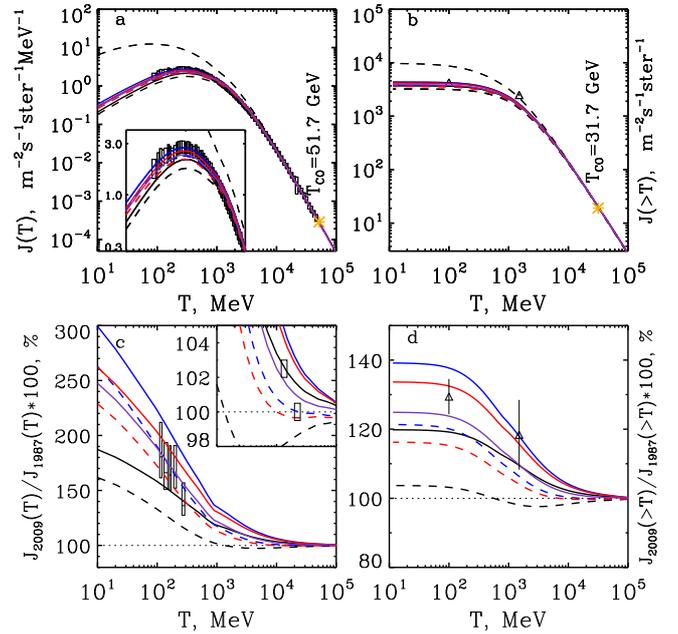}
  \caption{The observed and calculated energy spectra of the GCR protons near the Earth in the solar minimum 23/24. In the panels: (a) the absolute differential spectra with the black rectangles for PAMELA data and the enlarged maximum part of the spectra in the insert; (b) the absolute integral spectra with the black triangles for the RBM data; (c) the relative to 1987 differential spectra with the low energy black rectangles for the ratios of PAMELA (2009) and IMP8/GME (1987) data, the high energy black rectangles for the ratios of neutron monitor Moscow and Tsumeb data for 2009 and 1987 and the enlarged high energy part of the spectra in the insert; (d) the relative to 1987 integral spectra with the black triangles for the ratios of the RBM data for 2009 and 1987. The upper black dashed curves in the upper panels are for the unmodulated spectra. Lines of different color and style are for different sets of additional parameters (see text).
  }
  \label{fig3}
 \end{figure}

First we tried the usual values $r_{max}=125$ AU, $\alpha_R=1$  for the additional parameters. The energy spectra for this case are shown by the black solid lines in all panels of Fig.\ref{fig3}. As can be seen the calculated GCR intensity for this case is too low especially for the relative integral spectrum and the low energy particles. The change of only $\alpha_R=0.8$ (the black dashed curves) makes the description even worse, while the change of only $r_{max}=80$ AU (the blue solid curves) results in the too high differential and integral spectra (both absolute for 2009 and relative to 21/22 solar minimum).

The simultaneous change of both $r_{max}=80$ AU and $\alpha_R=0.8$ (the blue dashed curves) makes the description for the low energy differential spectra better, but the high energy relative differential spectrum and the integral spectrum lay much below the observed data. At last the calculations with $r_{max}=90$ AU, $\alpha_R=1$ (the red solid curves) look as acceptable, if we use the Moscow NM data as the high energy observations. On the other hand the differential spectra calculated with $r_{max}=90$ AU, $\alpha_R=0.8$ (the red dashed curves) are acceptable, if the Tsumeb NM data are used. However, the last case totaly contradicts to the RBM integral data. The intermediate case ($r_{max}=90$ AU, $\alpha_R=0.9$; the violet solid lines) is the best in general and is our choice for the calculations of the time and energy behavior of the GCR intensity around the 23/24 solar minimum in \cite{Krainev_etal_ICRC33_0305_2013}.

Note that the use of lower $\alpha_R$ (the dashed curves) resulted in the decrease of the high energy spectra in 2009 when compared with 2009 in conformity with the NM Tsumeb data. Nevertheless how to imitate properly the change of the HMF inhomogeneities spectrum reported in \cite{Grigoriev_Starodubtsev_BullRAS_Physics_75_801_2011,Gerasimova_etal_ICRC33_0266_2013} is still an open question.

Only the spectra for $A$--negative period (2009) are shown in Fig.\ref{fig3}. However, we calculated them also for the $A>0$ cases with the same parameters and the yellow stars shown in the upper panels of Fig.\ref{fig3} indicate the crossovers of the calculated spectra for the last case ($r_{max}=90$ AU, $\alpha_R=0.9$). For the differential energy spectra $T_{co}=51.7$ GeV (in contrast to $T_{co}=9.3$ GeV for the pair of solar minima 21/22 and 22/23). It means that if the heliospheric situation during the next solar minimum 24/25 ($\approx$ 2018) is the same as that in the minimum 23/24, then according to our calculations the maximum intensity for all cosmic ray detectors including the high energy ones (neutron monitors, the muon telescopes) will be higher in 2018 than in 2009, in contrast to what is observed in 1997 when the neutron monitors counted less high energy cosmic rays than in 1987.

\section{Conclusions}
\noindent 1. When we compare the last three solar minima we can isolated two group of the heliospheric factors important for the GCR modulation. The first group includes the main modulating factors: the strength of the heliospheric magnetic field, the tilt of the heliospheric current sheet and the solar wind velocity, all of these characteristics showing gradual change from solar minimum 21/22 through 22/23 to 23/24. The second group includes the additional heliospheric factors: the dynamic pressure of the solar wind and the spectral index of the heliospheric magnetic field inhomogeneities, these characteristics changing rather abruptly with the beginning of solar cycle 23.

\noindent 2. Using rather simple model it is possible to reproduce the main features of the space and energy behavior of the GCR intensity in the solar activity minima 21/22 and 22/23 using some set of the model parameters, the main modulating factors averaged for one year before the moments of maximum GCR intensity and the usual additional factors: the effective size of the modulation region $r_{max}=125$ AU and the index of the rigidity dependence of the diffusion coefficients $\alpha_R=1$.

\noindent 3. The most simple way to describe the unusually soft energy dependence of the maximum GCR intensity near the Earth in the last solar minimum 23/24 (2009) is to use the lower additional factors: $r_{max}=90$ AU and, probably, $\alpha_R=0.9$. This reduction of the modulation region and change of the rigidity dependence of the diffusion coefficient is substantiated by the observed degradation of the solar wind and getting more laminar heliospheric magnetic field during solar cycle 23.

\noindent 4. If the heliospheric situation during the next solar minimum 24/25 ($\approx$ 2018) is the same as that in the minimum 23/24 (except the polarity of the heliospheric magnetic field), then the maximum intensity for all cosmic ray detectors including the high energy ones (neutron monitors, the muon telescopes) will be higher in 2018 than in 2009.

\vspace*{0.5cm}
\footnotesize{{\bf Acknowledgment:}{ We thank the Russian Foundation for Basic Research (grants 11-02-00095a, 12-02-00215a, 13-02-00585a, 13-02-10006k) and the Program "Fundamental Properties of Matter and Astrophysics" of the Presidium of the Russian Academy of Sciences.}}


\begin{thebibliography}{}

\bibitem{Sheeley_ASPCS_428_3_2010}
N.R. Sheeley, Jr., ASP Conference Series 428 (2010) 3

\bibitem{Ahluwalia_Jackiewicz_ASR_50_662–668_2012}
H.S. Ahluwalia and J. Jackiewicz, Advances in Space Research 50 (2012) 662–668 doi:10.1016/j.asr.2011.04.023.

\bibitem{Svirzhevsky_etal_ICRC31_icrc1105_2009}
N.S. Svirzhevsky et al., Proc. 31st ICRC \verb| http://icrc2009.uni.lodz.pl/proc/pdf/icrc1105.pdf| (2009)

\bibitem{Krainev_Kalinin_ICRC31_icrc1043_2009}
M.B. Krainev, M.S. Kalinin, {\it Proc. 31st ICRC} \verb| http://icrc2009.uni.lodz.pl/proc/pdf/icrc1043.pdf| (2009)

\bibitem{McDonald_etal_GRL_37_L18101_2010}
F.B. McDonald et al., Geophys. Res. Lett. 37 (18) (2010) L18101 doi:10.1029/2010GL044218.

\bibitem{Mewaldt_etal_ApJL_723_L1-L6_2010}
R.A. Mewaldt et al., Astrophys. J. Lett. 723 (2010) L1-L6 doi:10.1088/2041-8205/723/1/L1.

\bibitem{Gushchina_etal_JP:CS_409_012169_2013}
R.T. Gushchina et al., Journal of Physics:Conference Series 409 (2013) 012169 doi:10.1088/1742-6596/409/1/012169.

\bibitem{Bazilevskaya_etal_ASR_49_784-790_2012}
G.A. Bazilevskaya et al., Advances in Space Research 49(4) (2012) 784-790 doi:10.1016/j.asr.2011.12.002.

\bibitem{Moraal_Stoker_JGR_115_A12109_2010}
H. Moraal and P.H. Stoker, JGR 115 (2010) A12109 doi:101029/2010JA015413.

\bibitem{Krainev_Kalinin_Bull_RAS_Physics_75_7862_2011}
M.B. Krainev and M.S. Kalinin, Bulletin of the Russian Academy of Sciences. Physics 75(6) (2011) 786–-789.

\bibitem{Gerasimova_etal_ICRC32_11_179}
S.K. Gerasimova etal, Proc. 32rd ICRC 11 (2011) 179-182.

\bibitem{Kota_Jokipii_ICRC32_11_12-14_2011}
J. Kota, J.R. Jokipii, Proc. 32rd ICRC 11 (2011) 12-14.

\bibitem{Bazilevskaya_etal_ICRC33_0274_2013}
G.A. Bazilevskaya et al., icrc2013-0274.pdf (2013)

\bibitem{Krainev_etal_ICRC33_0305_2013}
M.B. Krainev et al., icrc2013-0305.pdf (2013)

\bibitem{Bazilevskaya_Svirzhevskaya_SSR_85_431-521_1998}
G.A. Bazilevskaya, A.K. Svirzhevskaya, Space Science Reviews 85 (1998) 431-521.

\bibitem{Stozhkov_etal_Preprint_LPI_14_2007}
Yu.I. Stozhkov et al., Preprint 14 Lebedev Physical Institute, Moscow, 77 p (2007)

\bibitem{NM_Moscow}
\verb|http://helios.izmiran.rssi.ru/cosray/days.htm|

\bibitem{OMNI_Site}
\verb|ftp://omniweb.gsfc.nasa.gov/pub/data/omni|

\bibitem{WSO_Site}
\verb|http://wso.stanford.edu/|

\bibitem{bib:izmiran} \verb|http://cr0.izmiran.rssi.ru/common/links.htm|

\bibitem{Mewaldt_SSR_online_2013}
R.A. Mewaldt, Space Science Reviews (2013) doi 10.1007/s11214-012-9922-0.

\bibitem{Grigoriev_Starodubtsev_BullRAS_Physics_75_801_2011}
V.G. Grigoriev and S.A. Starodubtsev, Bulletin of the Russian Academy of Sciences. Physics 75 (2011) 801–804.

\bibitem{Gerasimova_etal_ICRC33_0266_2013}
S.K. Gerasimova et al., icrc2013-0266.pdf (2013)

\bibitem{Parker_PhysRev_110_1445_1958}
E.N. Parker, Phys. Rev. 110 (1958) 1445.

\bibitem{Krymsky_GaA_4_977_1964}
G.F. Krymsky, Geomagnetism and Aeronomy 4 (1964) 977-985 (in Russian).

\bibitem{Jokipii_Levy_Hubbard_ApJ_213_861_1977}
J.R. Jokipii et al., Astrophysical Journal 213 (1977) 861-861.

\bibitem{Jokipii_Kota_GRL_16_1_1989}
J.R. Jokipii and J. K\'ota, GRL 16 (1989) 1-4.

\bibitem{Krainev_etal_JP:CS_409_012016_2013}
M.B. Krainev et al., Journal of Physics:Conference Series 409 (2013) 012016 doi:10.1088/1742-6596/409/1/012016.

\bibitem{Potgieter_etal_SpaceSciRev_97_295_2001}
M.S. Potgieter et al., Space Sci. Rev. 97 (2001) 295–307.

\bibitem{Krainev_Kalinin_AIPCP_1216_371-374_2010}
M.B. Krainev and M.S. Kalinin, AIP Conf. Proc. 1216 (2010) 371-374.

\bibitem{Langer_Potgieter_JGR_109_1-12_2004}
U.W. Langer, M.S. Potgieter, JGR 109 (2004) 1-12.

\bibitem{Potgieter_etal_Apxiv_1302.1284_2013}
M.S. Potgieter et al., Apxiv-1302.1284.pdf (2013)

\bibitem{Adriani_etal_ApJ_765_2_91_2013}
O. Adriani et al., ApJ 765 (2013) 91.

\bibitem{McDonald_SpaceScience_ser_ISSI_33_1998}
F.B. McDonald, in Cosmic Rays in the Heliosphere (eds.) Fisk L A et al. Space Science series of ISSI (1998) 33-50.

\end{thebibliography}
\end{document}